\documentclass{article}

\usepackage{arxiv}

\usepackage[utf8]{inputenc} % allow utf-8 input
\usepackage[T1]{fontenc}    % use 8-bit T1 fonts
\usepackage{hyperref}       % hyperlinks
\usepackage{url}            % simple URL typesetting
\usepackage{booktabs}       % professional-quality tables
\usepackage{amsfonts}       % blackboard math symbols
\usepackage{nicefrac}       % compact symbols for 1/2, etc.
\usepackage{microtype}      % microtypography

\usepackage{amsmath}
\usepackage{bbm}
\usepackage{amssymb,amsfonts}
\usepackage{xcolor}

\usepackage{caption}
\usepackage{subcaption}

\usepackage{booktabs}
\usepackage{multirow}
\usepackage{graphicx}

\title{Challenges in Forecasting Malicious Events from Incomplete Data}

\author{
    Nazgol Tavabi\thanks{Both authors contributed equally to this research.}\\
    University of Southern California\\ 
    Information Sciences Institute\\
    4676 Admiralty Way\\
    \texttt{nazgolta@isi.edu}
  \And
    Andr\'es Abeliuk$^*$\\
    University of Southern California\\ 
    Information Sciences Institute\\
    4676 Admiralty Way\\
    \texttt{aabeliuk@isi.edu}
  \And
    Negar Mokhberian\\
    University of Southern California\\ 
    Information Sciences Institute\\
    4676 Admiralty Way\\
    \texttt{nmokhber@isi.edu}
  \And
    Jeremy Abramson\\
    University of Southern California\\ 
    Information Sciences Institute\\
    4676 Admiralty Way\\
    \texttt{abramson@isi.edu}
 \And
    Kristina Lerman\\
    University of Southern California\\ 
    Information Sciences Institute\\
    4676 Admiralty Way\\
    \texttt{lerman@isi.edu}
}

\begin{document}
\maketitle

\begin{abstract}
% The ability to accurately predict cyber-attacks would enable organizations to mitigate their growing threat and avert the financial losses and disruptions they cause. How predictable are cyber-attacks? Researchers have attempted to combine external data---from vulnerability disclosures to discussions on Twitter and the darkweb---with machine learning algorithms to learn indicators of impending cyber-attacks. One limitation of machine learning methods is that they require training data, in this case records of successful attacks, to learn predictive patterns. However, appropriate training data is often incomplete and highly filtered: e.g., successful cyber-attack represent just a tiny fraction of all attempted attacks, with the vast majority of which are stopped by the security appliances deployed by the target. As we show in this paper, the process of filtering itself reduces the predictability of cyber-attacks. As a result, the small number of attacks that do penetrate the target's defenses may not create useful training data for machine learning algorithms to learn predictive models. We empirically quantify the loss of predictability due to filtering using data from two organizations. Our work identifies limits to forecasting cyber-attacks from highly filtered data. %This logic does not apply to attempted attacks, however, which should be predictable. 

The ability to accurately predict cyber-attacks would enable organizations to mitigate their growing threat and avert the financial losses and disruptions they cause. But how predictable are cyber-attacks? Researchers have attempted to combine external data -- ranging from vulnerability disclosures to discussions on Twitter and the darkweb -- with machine learning algorithms to learn indicators of impending cyber-attacks.
% A limitation of machine learning methods in this context is that they require training data -- e.g. successful attacks -- to learn predictive patterns. However, training data is often limited to just the 
However, \textit{successful} cyber-attacks represent a tiny fraction of all \textit{attempted} attacks: the vast majority are stopped, or filtered by the security appliances deployed at the target. As we show in this paper, the process of filtering reduces the predictability of cyber-attacks. %As a result, the small number of attacks that do penetrate the target's defenses may not create useful data for machine learning algorithms to learn predictive models. 
The small number of attacks that do penetrate the target's defenses follow a different generative process compared to the whole data which is much harder to learn for predictive models. This could be caused by the fact that the resulting time series also depends on the filtering process in addition to all the different factors that the original time series depended on. We empirically quantify the loss of predictability due to filtering using real-world data from two organizations. Our work identifies the limits to forecasting cyber-attacks from highly filtered data. %This logic does not apply to attempted attacks, however, which should be predictable. 
\end{abstract}


\maketitle

\section{Introduction}
Malicious behavior is increasingly common in online life. Social media platforms are racing to develop tools to detect---and in some cases anticipate---malicious behaviors in the form of manipulation, deception, misinformation, and cyberbullying.  
Cybercrime is one example of malicious behavior that has resulted in large financial losses, political and security risks. The 2016 hacking of the Democratic National Committee server was arguably a turning point in the 2016 US presidential elections. The leaks of  potentially embarrassing emails upended the race and upset existing polls.
%Could such an attack been foreseen? 
%In addition, cyber-threats come from a variety of sources, including foreign nations, terrorist and criminal organizations, activist groups (aka ``hacktivists''), and individual hackers. 
The ability to anticipate cybercrime --- and other security threats, such as violent protests in a country --- would allow organizations  to mitigate the risks associated with such disruptions. As an age-old saying goes: ``to be forewarned is to be  forearmed.''

How predictable is malicious behavior? %Answering this question gives us insights into the ability to predict other types of malicious behaviors, since cybercrime predictions uses some of the same methods. 
To narrow the scope of the question, we consider the predictability of cybercrime, since forecasting cybercrime shares many challenges of forecasting other types of malicious activities. 
In the case of cybercrime, conventional wisdom says indicators often exist that can be leveraged for detection and prediction. Successfully executing a cyber-attack requires preparation and planning: hackers carry out reconnaissance about the potential targets, identify its vulnerabilities, acquire relevant tools and exploits, etc. All these activities leave traces within the openly available data (as well as within the patterns of attacks themselves) that allow for forecasting new cyber-attacks. 
 
\begin{center}
\begin{figure}[t]
\centering
\includegraphics[width=0.75\columnwidth]{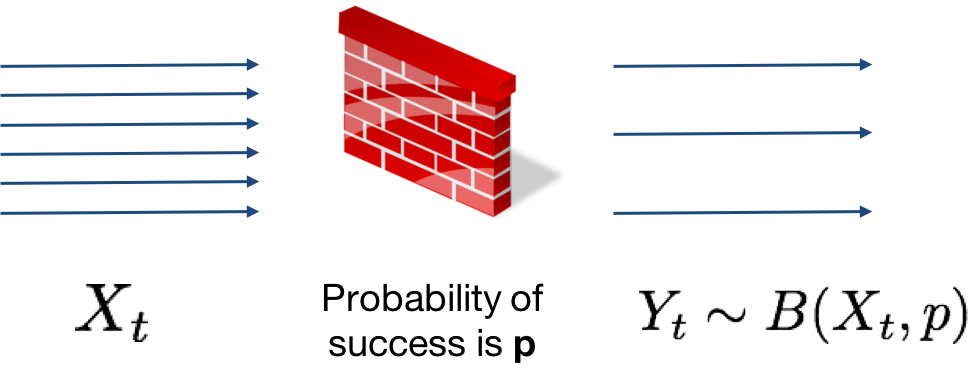}
\caption{Sketch of the filtering framework. In the context of cyberattacks, $X$  is the number of attempted attacks per day;  $Y$ is the number of successful attacks per day observed by the target.  The probability of an attack being successful is $p$. The Binomial distribution $B(n,p)$ is used to model the time series $Y$. }
\label{fig:sketch}
\end{figure}
\end{center}

%\paragraph{Our contribution:} 
An important consideration for an AI-based solution to cyber-attack prediction is that the successful attacks used to train machine learning models make up a \textit{small fraction} of all attacks. In reality, the vast majority of attacks are stopped by the target's defenses: firewalls, domain blockers, spam filters, etc, as illustrated in Figure~\ref{fig:sketch}. Only a small fraction of attempted attacks reach the victim and are recorded as training data for machine learning algorithms.  
This also pertains to other types of malicious behaviors, which are at best only partially observed, as malicious actors attempt to obfuscate their behaviors.
As we show in this paper, %the mere act of this filtering 
having access to only a subset of malicious events for use in training 
reduces the predictive utility of the data, and accuracy of the models learned from it. In this paper, we demonstrate this important problem in the context of cyber-attack forecasting.

This work makes the following two contributions: 1) We quantify the impact of incomplete observation due to filtering on the predictability of malicious events using real-world email data from two distinct organizations and 2) We show that predictability decreases and prediction error grows when more malicious emails are filtered by the organization's defenses. 
%Our work shows that  predicting cyber-attacks with AI models learned from data is more complex than originally thought, and practitioners should strive to get as unfiltered a sample of ground truth data about cyber-attacks as they possibly can to improve the predictive utility of the AI models.
Our work identifies an important challenge researchers have to consider when forecasting malicious activity, including cyber-attacks.

\section{Related Work}
%\section{Related Work}
%The work presented here covers a wide cross-section of active research areas.
%We focus on %literature survey on 
%machine learning methods for cyber-attack forecasting, as well as methods for using external signals to enhance prediction capabilities.

%\paragraph{Cyber Indicent Forecasting and Assessment}
Researchers have used state-of-the-art machine learning methods to forecast cyber-attacks and extract predictive indicators from available data sources. 
Works such as \cite{werner2017time,zhan2015predicting,okutan2017predicting,goyal2018discovering} have trained models to find associations between such indicators and successful attacks.
Approaches using temporal forecasting include prediction of cyber breaches \cite{Xu.2018.Modeling-and-Pr}.
This type of analysis models the inter-arrival time of breach incidents inter-arrival as a stochastic process, described by an auto-regressive and Moving Average (ARMA).
The authors also show a decrease between event inter-arrival times, indicating that cyber-attacks are becoming more frequent.
A Bayesian framework for cyber-attack prediction is presented in \cite{Wu.2012.Cyber-Attacks-P}.  The authors use attack graphs to represent and enumerate possible system vulnerabilities and attack paths.
While this approach is promising, in practice it may not be actionable, as it requires an accurate picture of all system attack surfaces and vulnerabilities, which is unrealistic in a modern enterprise network.
The authors in \cite{Allodi.2017.Security-Events} propose a methodology to determine enterprise-level risk assessment against "untargeted" attacks, testing the framework on data accrued from a large financial institution.
The work in \cite{Bakdash.2018.Malware-in-the-} presents a Bayesian State Space Model (BSSM) model for forecasting cyber-attack events, and can do so with reasonable accuracy for non-bursty events up to one week out.  This result is promising, however, the data set itself is exactly the sort of filtered view that this work addresses.
The training data used in the study includes approximately seven years of weekly analyst-verified cyber incidents at a large US Department of Defense enterprise.
%The results in the paper are promising, although the data set used is itself highly filtered.
%It does not consider either attacks that were blocked, nor does it consider successful attacks that went undetected by the enterprise systems and human analysts.    
%\paragraph{External Signals}

Another line of research is to use a variety of external data to improve predictions. This includes
\cite{Ritter.2015.Weakly-Supervis} and \cite{Sapienza.2018.DISCOVER}, which uses external data from Twitter discussions to automatically learn keywords associated with emerging cyber threats, such as malware or botnet names.
Other works identify patterns within discussions of vulnerabilities on darkweb~\cite{almukaynizi2018darkmention} or sentiment of posts in hacker forums~\cite{deb2018predicting},  predictive of future cyber-attacks,
or identify software vulnerabilities that are likely to get exploited \cite{sabottke2015vulnerability,tavabi2018darkembed}. 
All of these approaches are useful in enhancing predictions of a specific domain based on external signals relevant to that domain, but none address the fundamental impact of data on the predictability of cyber-attacks.

\section{Limits of Predictability}
Our work builds upon our recent discovery of a fundamental limit to the predictability of partially observed dynamic systems. We developed a mathematical framework to characterize the loss of information in sampled time series~\cite{abeliuk2019predicting}. The framework allows us to quantify how the predictability of dynamic systems is affected by temporal sampling under a variety of sampling conditions. In this work, we will extend our framework to quantify the impact of predictability loss on real-world forecasting tasks of interest in security applications. Specifically, we will measure how partial observation affects the accuracy of forecasting cyberattacks. %and political violence and protest around the world, among other applications.

%\paragraph{Partially Observed Dynamic Systems.}
Consider a dynamic process generating events, (e.g.) cyberattacks against an organization or protests in a region.  We represent the time series of this \emph{ground truth} data as $X=[X_1,X_2,\ldots,X_T]$, each entry representing the number of events at time $t$. % We refer to this time series as the \textit{ground truth signal}.
Observers of this process may not see all events. Twitter, for example, makes only a small fraction ($\leq10\%$) of messages posted on its platform programmatically available, and an organization's defenses may stop the vast fraction of attempted attacks from reaching end users. We refer to the time series of the \emph{observed} data as $Y=[Y_1,Y_2,\ldots,Y_T]$.
Figure \ref{fig:sketch} depicts the cyber-attack prediction problem  under our framework.
% Next, we formalize how both time series are related. a stylized version of

%\paragraph{Predictability in Sampled Time Series.}
We model partial observability as a stochastic sampling process, where each event has some probability $p$ to be observed, independent of other events.
Therefore, we can represent the observed data as:
$Y_t \sim B([X_t],p), $
where $B([X_t],p)$ represents the Binomial distribution with parameters $[X_t]$ for sample size and success probability $p$. 
% The corresponding information preservation (success) rate is $p=1-q$.
% Events are often driven by some external factors. For example, the price of Bitcoin may make some hacking tools cheaper to acquire, increasing the likelihood of cyberattacks, and news reports may drive social media posts on specific topics.
% These external factors may help predict the ground truth signal. We refer to the time series of such informative external factors as the \textit{external signal} $S=S_1,S_2,\ldots,S_T$.

Under this framework, we derived the following theoretical result % two theoretical results
 ~\cite{abeliuk2019predicting}:
\begin{description}
\item[Decay of auto-correlation of the observed signal.]
The auto-correlation of the observed signal $Y$ decays monotonically at lower sampling rates. (lower probability $p$)
% \begin{equation}
% \label{eq:autocorr}
%     \rho_{Y_i,Y_j} \approx \frac{ p^2 \mathrm{Cov}(X_i,X_j)}{p^2\mathrm{Var}(X) + p(1-p)\E[X]}.
% \end{equation}
%  When $p=1$ (i.e., in presence of complete observation), we recover the auto-correlation of the ground truth signal $X$. At lower sampling rates, the auto-correlation decays as postulated above.
%\item[Decay of covariance with the external signal.]
%The correlation between the observed and external signals degrades linearly at lower sampling rates .
\end{description}

\paragraph{From theory to application.}
In this paper, we aim to apply the theoretical framework -- describing the effects of sampling on predictability -- to real-world situations in the context of predicting cyber-attacks. We will explore whether the simplified assumptions of the theoretical model stand up to the test of empirical verification. Thus, our current study addresses two research questions:
\begin{description}
\item[RQ1]{Our past work shows a linear (autocorrelation) loss of predictability due to sampling. How does sampling affect non-linear forecasting methods? Are these methods more/less robust than linear ones?}
\item[RQ2]{For analytical purposes, partial observability is modeled as a random sampling, however, this is rarely the case in reality. Does incomplete data affect real-world cyber-attack forecasting scenarios?  }
\end{description}

Our past work primarily focused on linear forecasting models, such as ARIMA, which we linked to linear measures of predictability, like autocorrelation. However, it is possible that due to nonlinear interactions, predictability may not be lost as quickly during sampling. %, but this is not captured by the linear predictability measures and models. explore this idea,
To test this theory, in this work we  investigate the possibility of mitigating the loss of information using non-linear relationships in data and metrics.
We will identify such cases and compensate for the loss of linear predictability using \emph{nonlinear forecasting models} and nonlinear measures of predictability, such as permutation entropy~\cite{Bandt2002permutation}.
We will use neural networks in the prediction tasks and compare their performance to linear forecasting models, such as ARIMA. We will identify applications where linear and non-linear predictability measures diverge and explore whether non-linear models can compensate for the loss of information. 

Second, the framework assumes that partial observability can be modeled as a Binomial sampling with a fixed probability of success. However, depending on the application, this may rarely be the case. For example, a Twitter sample obtained through their API may not be a good representation of the data due to non-homogeneous sampling  \cite{morstatter2013sample}. In malicious email detection, multiple layers of spam filter techniques are combined for advanced protection~\cite{shandilya2014amulti}. In this work, we recreate as close as possible the sampling process over the attempted attacks by sequentially turning on these multiple layers of protection.

\begin{figure}[t]
\centering
  \begin{subfigure}[b]{0.48\textwidth}
  \centering
    \includegraphics[width=\textwidth]{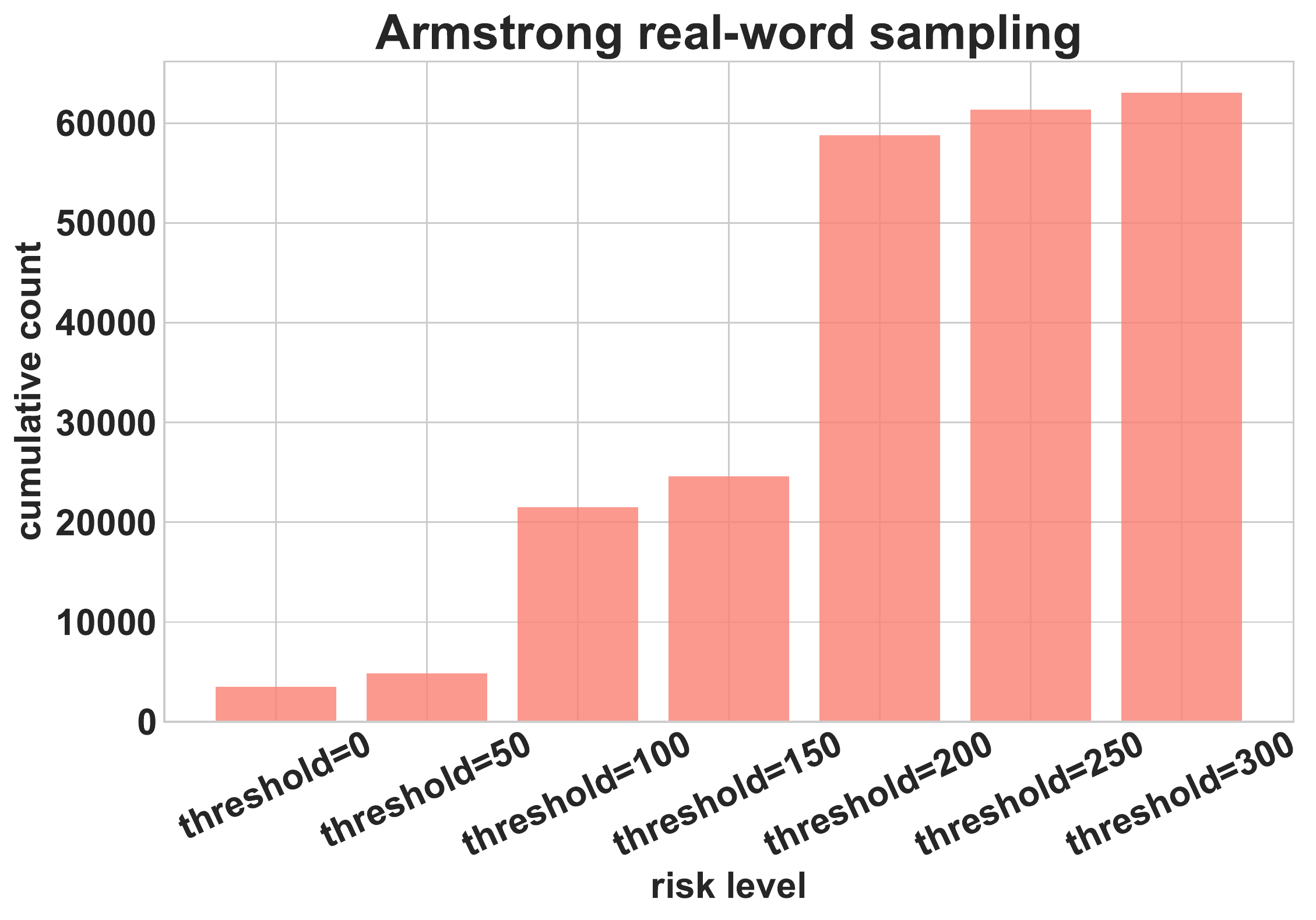}
    \caption{Number of threat messages in Armstrong data based on their risk score threshold}
    \label{fig:dist_arm}
  \end{subfigure}%
   \hspace{1em}%
 \centering
  \begin{subfigure}[b]{0.48\textwidth}
  \centering
    \includegraphics[width=\textwidth]{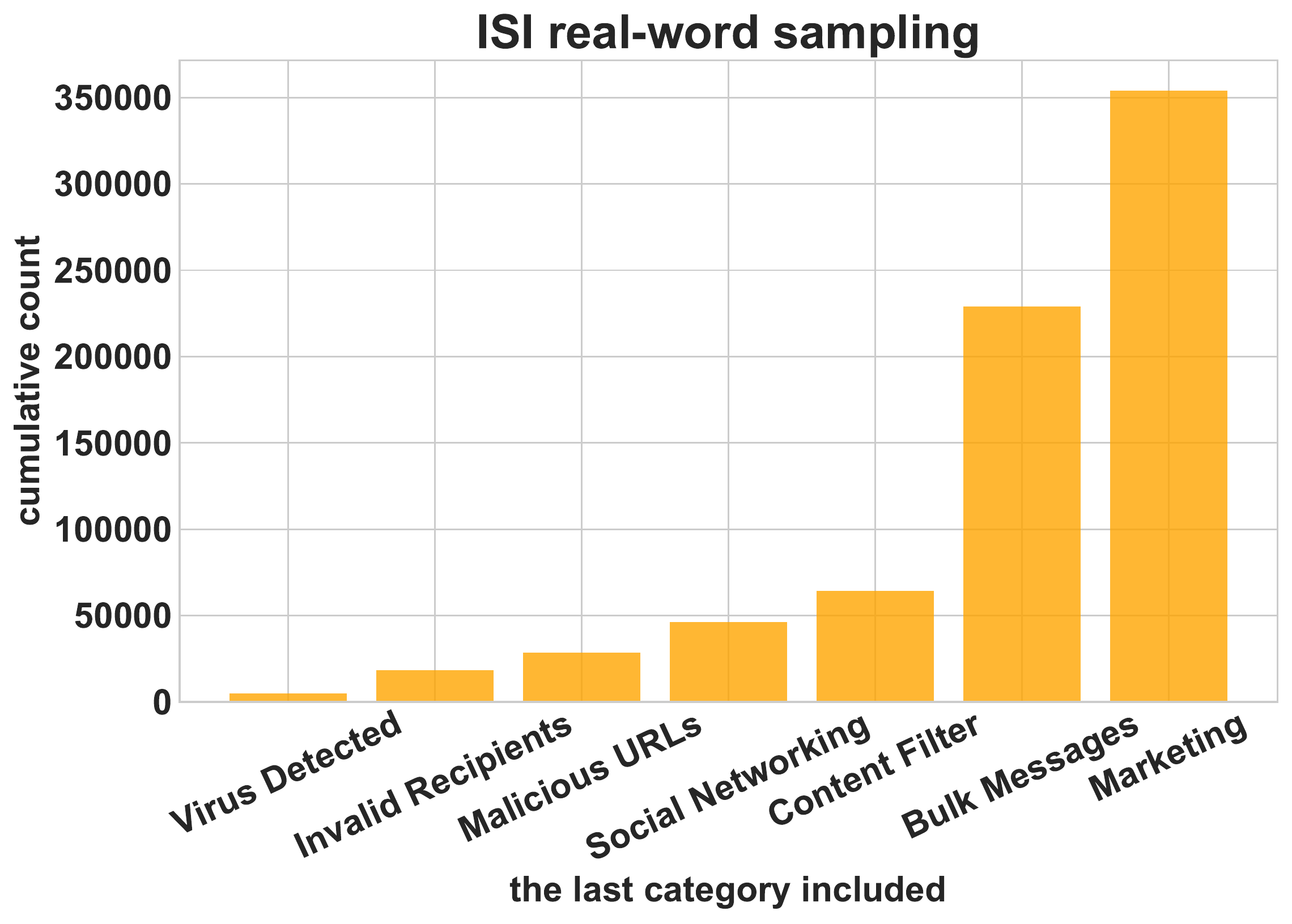}
    \caption{Cumulative number of messages in each category within ISI data.}
    \label{fig:barplot_isi}
  \end{subfigure}
  \caption{Armstrong and ISI malicious emails data. }
\end{figure}

\section{Methods}

\subsection{Data}
In this work, we explore email metadata sourced from two distinct enterprises.  
The first, denoted \textit{ISI}, represents email metadata from the University of Southern California's Information Sciences Institute.
The second is an anonymized applied science and technology company we denote as \textit{Armstrong}.
In both instances, we analyze the output from the specific enterprise's spam appliance, which does a number of classification and filtering steps in order to protect end-users from potentially malicious emails. %\cite{sculley2008filtering}.
We explain these data sets in more detail in the following section.

\subsubsection{ISI}
%\andres{Negar include a figure showing the data; and also give some simple descriptive statistics like (daily mean, variance, min max...)}
% \paragraph{ACLED?}
The ISI data set consists of daily summaries of all incoming mail activity from August 2018 through July 2019. 
These summaries contain count data for several threat classifications, as determined by the ISI spam appliance.
%(see Figure \ref{fig:isi-random}).
 The average number of total daily malicious emails is 2233 with a minimum of 0, maximum of 7163 and the standard deviation of 1733. The specific message classifications are as follows:

\begin{itemize}
  %\item \textit{Reputation Filter}: Messages originating from known malicious ("blacklisted") domains or severs 
  \item \textit{Invalid Recipients}: Messages dropped by a Lightweight Directory Access Protocol (LDAP) accept-query process
  \item \textit{Virus Detected}: Messages with a malicious payload, as detected via the appliance virus scanner
  %\item \textit{Spam Detected}: Messages over the spam heuristic threshold 
  \item \textit{Messages with Malicious URLs}: Messages containing malicious URLs, as determined via appliance blacklists and heuristics, in the message body or attachments
  \item \textit{Stopped by Content Filter}: Messages having contents related to gaming, pornography, weapons, etc.
  \item \textit{Marketing}: "Gray"-mail marketing messages sent by recognized marketing groups
  \item \textit{Social Networking}: Messages from social networks, forums, etc.
  \item \textit{Bulk}: Advertising and marketing messages sent by unrecognized sources.
\end{itemize}

%This data was collected from a Web security appliance deployed in the email infrastructure of an academic institution (USC/ISI). 
%The security software detects and filters potential malicious incoming e-mails based on policy controls. 
%ISI data was collected from August 2018 until July 2017. %and receives ... \nazgol{malicious} emails on average each day. 
%\andres{the following seems redundant give the above?}
Security appliances of this company classify %{unwanted} 
unsolicited emails by categories and block any/all categories based on the organization's predefined policy. %List of these categories is given below:   %We collected over ... months worth of daily threat emails detected by different filters. %To make the analogy to our framework, we treat each filter as a sample of the total number of attempted attacks. Figure \ref{fig:exp2} depicts the time series for each filter, as well as the total number of threat messages.

% \usepackage{multirow}
% \usepackage{graphicx}
%\begin{table}[]
%\begin{tabular}{|l|l|l|}
%\hline
%\textbf{Data Set}                  & \textbf{Armstrong}                  %            & \textbf{ISI}     \\ \hline
%\hline
%{\textbf{Features}} & impostorScore                                   & Reputation       \\ \cline{2-3} 
%                                   & malwareScore                        %            & Spam             \\ \cline{2-3} 
%                                   & spamScore                           %            & Malware          \\ \cline{2-3} 
%                                   & phishScore                          %            & Content Filtered \\ \cline{2-3} 
%                                   & classification  & Greymail         \\ \cline{2-3} 
%                                   & malice type                       & Virus            \\ \hline
%\end{tabular}%
%\caption{Summary of features from email data sets.  Classification is one of [spam/phishing/impostor] and malice type is one of [URL/attachment].  }
%\end{table}

\begin{figure*}[h!]
\centering
\includegraphics[width=\linewidth]{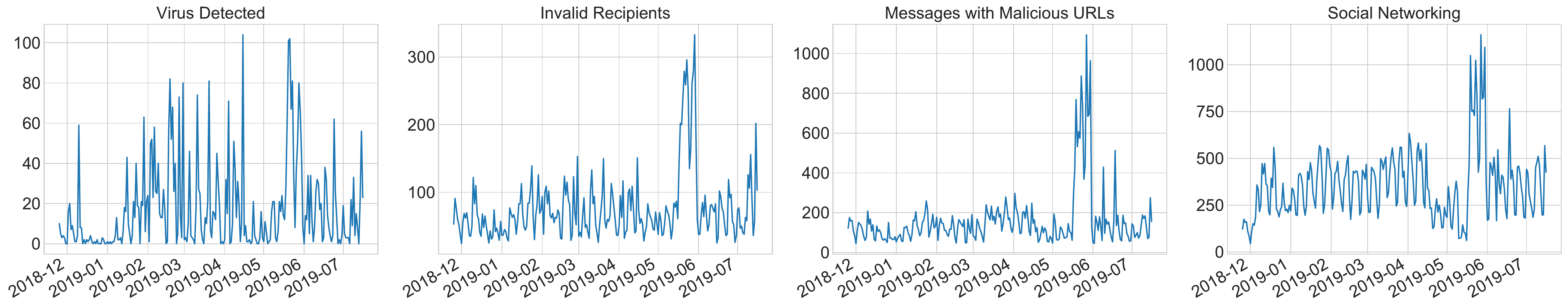}
\includegraphics[width=0.75\linewidth]{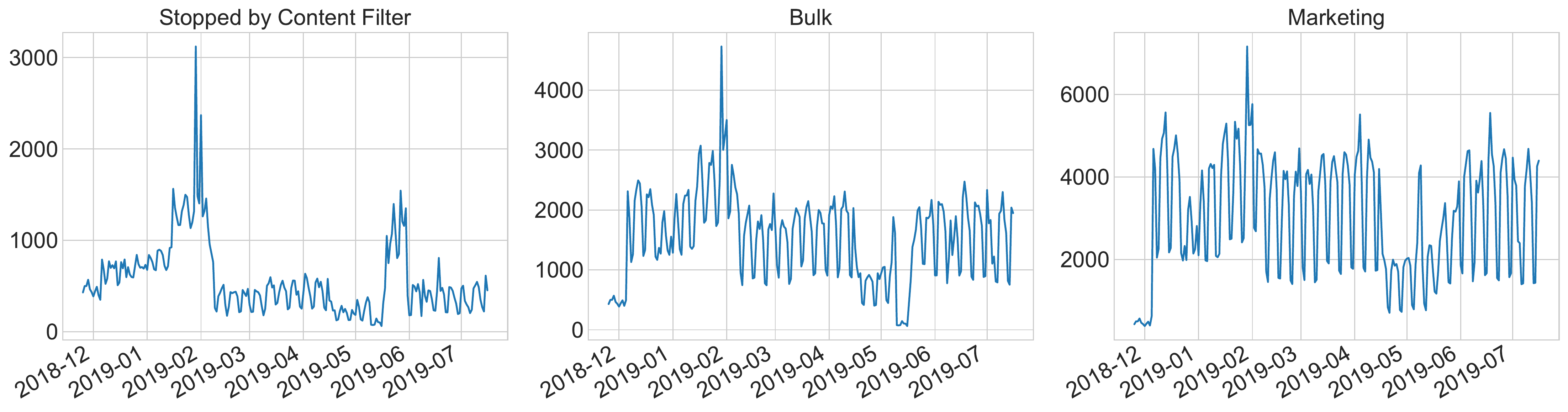}
\caption{Cumulative time series of categories in ISI data used for real-world sampling. The top leftmost plot is the time series of messages from Virus Detected category only, we add other categories to sampled data one-by-one. The last plot (titled as Marketing) contains messages from all malicious categories.}
\label{fig:isi_ts}
\end{figure*}

\subsubsection{Armstrong}
%\andres{Negar include a figure showing the data; and also give some simple descriptive statistics like (daily mean, variance, min max...)}
%\nazgol{Jeremy can you add descriptions for these scores and ISI scores plus general description of the both datasets?}
This data was collected from a web security appliance deployed in the email infrastructure of the Armstrong organization. Armstrong data ranges from February 2018 to January 2019 and has on average 184 email messages per day. The minimum number of daily attacks is 0 and the maximum number is 1094 with the standard deviation is 163. This security appliance assigns 4 scores to each email, described below:
\begin{itemize}
  \item \textit{Impostor Score}: An aggregate score of rules and heuristics that compare (e.g.) message credentials and metadata to message contents in order to determine if the sender is who they purport to be
  \item \textit{Malware Score}: A determination of confidence that the message contains a malicious attachment or URL 
  \item \textit{Spam Score}: A score based on spam heuristics such as message keywords, metadata agreement, and other traditional methods
  \item \textit{Phish Score}: An indication of how confident the email appliance is that the email is attempting to elicit information from the recipient maliciously 
\end{itemize}  
  %The aforementioned scores \textbf{could be used/ are used ?} as filtering mechanism to prevent cyberattacks. %However based on the thresholds chosen for filtering, predicting the attacks that are actually able to pass the filters becomes more difficult.

\subsection{Sampling}\label{sec:sampling}
To understand the effects of sampling on predictability, we used both random sampling and the more realistic real-world sampling.

\subsubsection{Random Sampling}
%\textbf{Negar to describe}
 For random sampling, we filter events uniformly at random with some probability. This sampling can be modeled as a Bernoulli trial with parameter $p$. While changing the parameter from $0$ to $1$ we change the sampling rate. Also, since this process is stochastic for each probability $p$ we sample the time series 50 times.

%In reality the sampling is not uniform, but dictated by the detection threshold of the security appliances. We model this more realistic sampling by filtering out emails that have been flagged by the security appliance in each category.

%As described in the section Data,
% TODO what is the best way to refer to section data?
 %for ISI data we have Invalid Recipients, Virus Detected, Messages with Malicious URLs, Stopped by Content Filter, Marketing, Social Networking, and Bulk (sorted by the number of total activities) as the seven categories of malicious activities. We consider the category with fewer activities more malicious than the ones with more activities. Therefore our real-world sampling starts with considering only the counts of Bulk messages received daily, and adding the number of daily activities in more risky categories one-by-one in seven steps.
 
 %In Armstrong data we have four scores of Phishing, Spam, Malware, and Imposter for each activity. For simulating the real-world sampling we set different thresholds on the summation of these scores. We start with the lower thresholds and then expand the amount of sampled data with increasing the threshold and including the more risky activities.
%\textbf{Describe the prediction pipeline}

\subsubsection{Real-world sampling}
 Real-world sampling for each dataset is described below.
\paragraph{ISI}
%The overall number of events detected in each category is shown in Figure~\ref{fig:isi-data}. 
To sample ISI data, we mimic the filtering done by the security appliances. We treat each category as a filter that prevents emails of that type to pass. %\note{The categories are different from the figure.}  
The order of filters is based on the number of emails in each group: we filtered out the group with the highest number of emails ("Marketing") first, %"Marketing", afterwards 
then "Bulk" and so on.  We chose this ordering as it roughly aligns with our data; we consider the maliciousness of the message to be inversely proportional to the frequency such message types are observed (i.e. spam is common, therefore not as dangerous).  In Figure~\ref{fig:barplot_isi}, we can see that more serious attacks like messages with viruses occur less frequently in comparison to marketing or bulk messages.
%We did not use data from categories "Reputation Filters" and "Spam Detected" because of the amount of data in these groups. "Reputation Filters" if included, would have contained 97\% of the data and "Spam Detected" would have contained 58\% of the remaining data, hence they would not have been good choices for sampling.
In Figure~\ref{fig:isi_ts} we can see the daily time series of counts for each threat categories. %remaining

%\begin{figure*}[t]
%\centering
%\includegraphics[width=0.7\linewidth]{Figures/isi_attacks.pdf}
%\caption{ Daily malicious activity detected by different security filters. The top left plot shows the total daily number of incoming messages with a security threat. }
%\label{fig:isi-data}
%\end{figure*}

\paragraph{Armstrong}
%\andres{this paragraph needs attention}
  We sum over the aforementioned threat scores, and denote this the Risk score.  This aggregate score has a potential range of $0$ to $400$ -- although it only ranges from $0$ to $300$ in the data -- is then used as a filter to sample emails. 
  The distribution of the risk scores in the Armstrong data is shown in Figure \ref{fig:dist_arm}.
  We filter the data by setting the risk score thresholds to $0, 50,\cdots 300$.  For example, threshold 0 allows only email messages with a 0 risk score to pass through the filter, and threshold 300 allows all emails to pass.

\begin{figure*}[h!]
\centering
\includegraphics[width=0.3\linewidth]{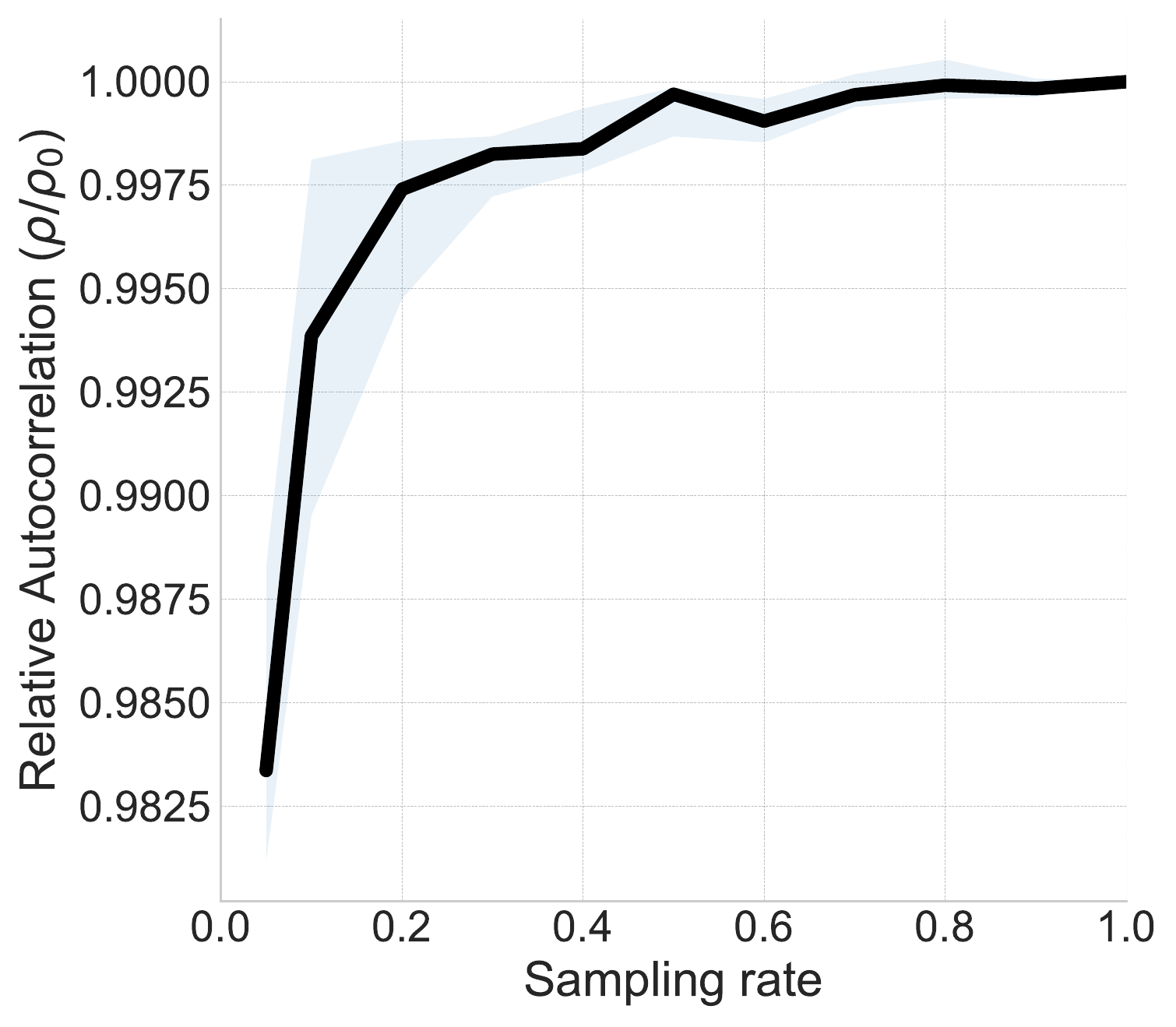}
\includegraphics[width=0.3\linewidth]{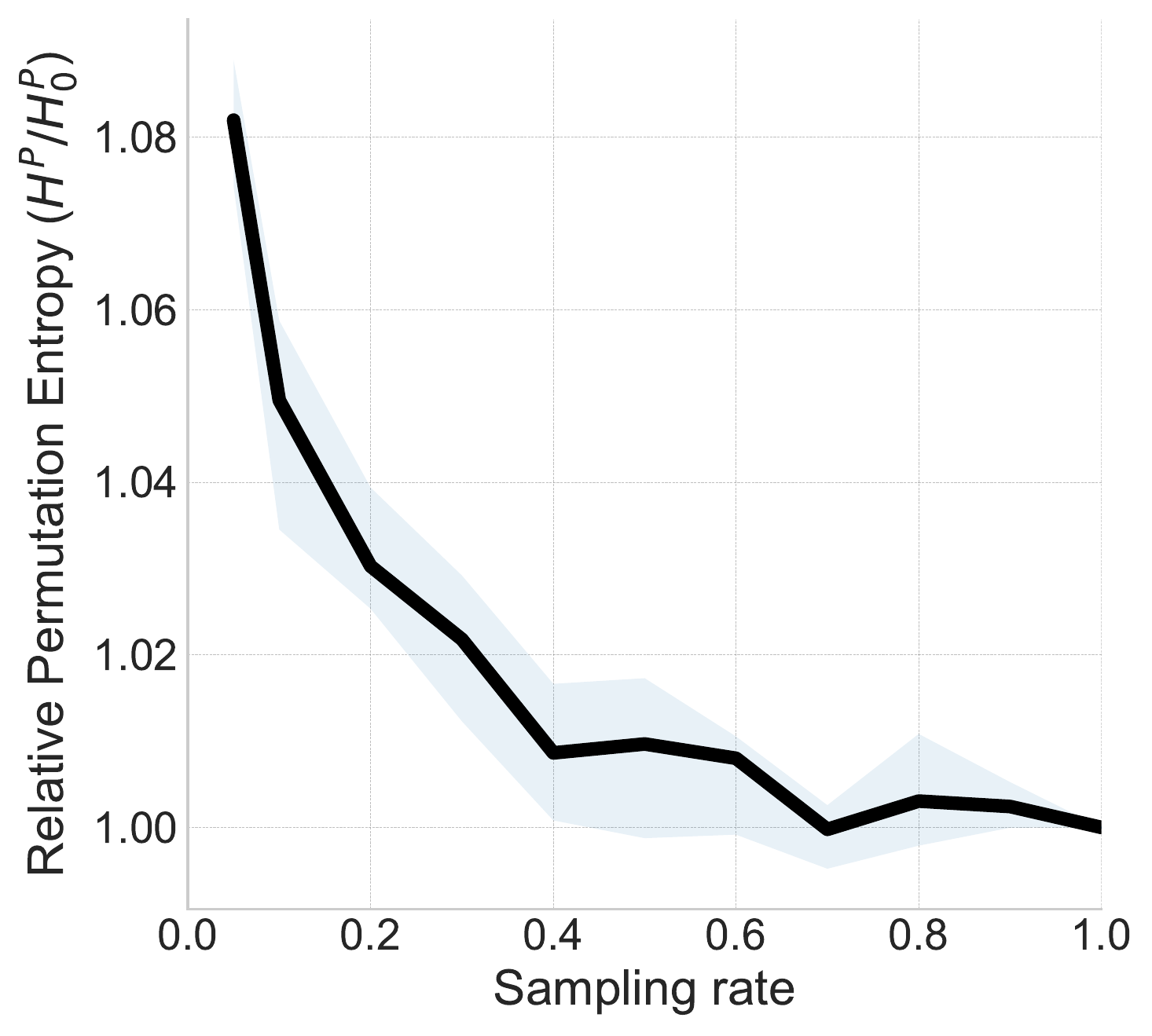}
\includegraphics[width=0.27\linewidth]{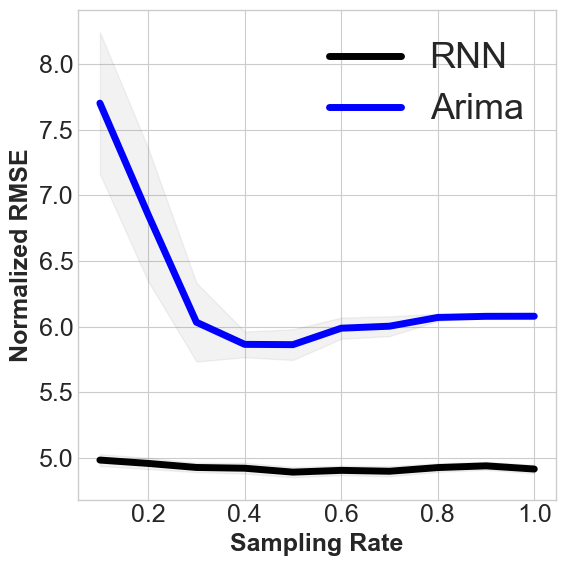}
\caption{Decay of predictability of (randomly) sampled ISI data. The (a) auto-correlation decreases at low sampling rates and (b) permutation entropy increases; and (c) and error of model-bases techniques increases}
\label{fig:isi-random}
\end{figure*}

\begin{center}
\begin{figure*}[t]
\centering
\includegraphics[width=0.3\linewidth]{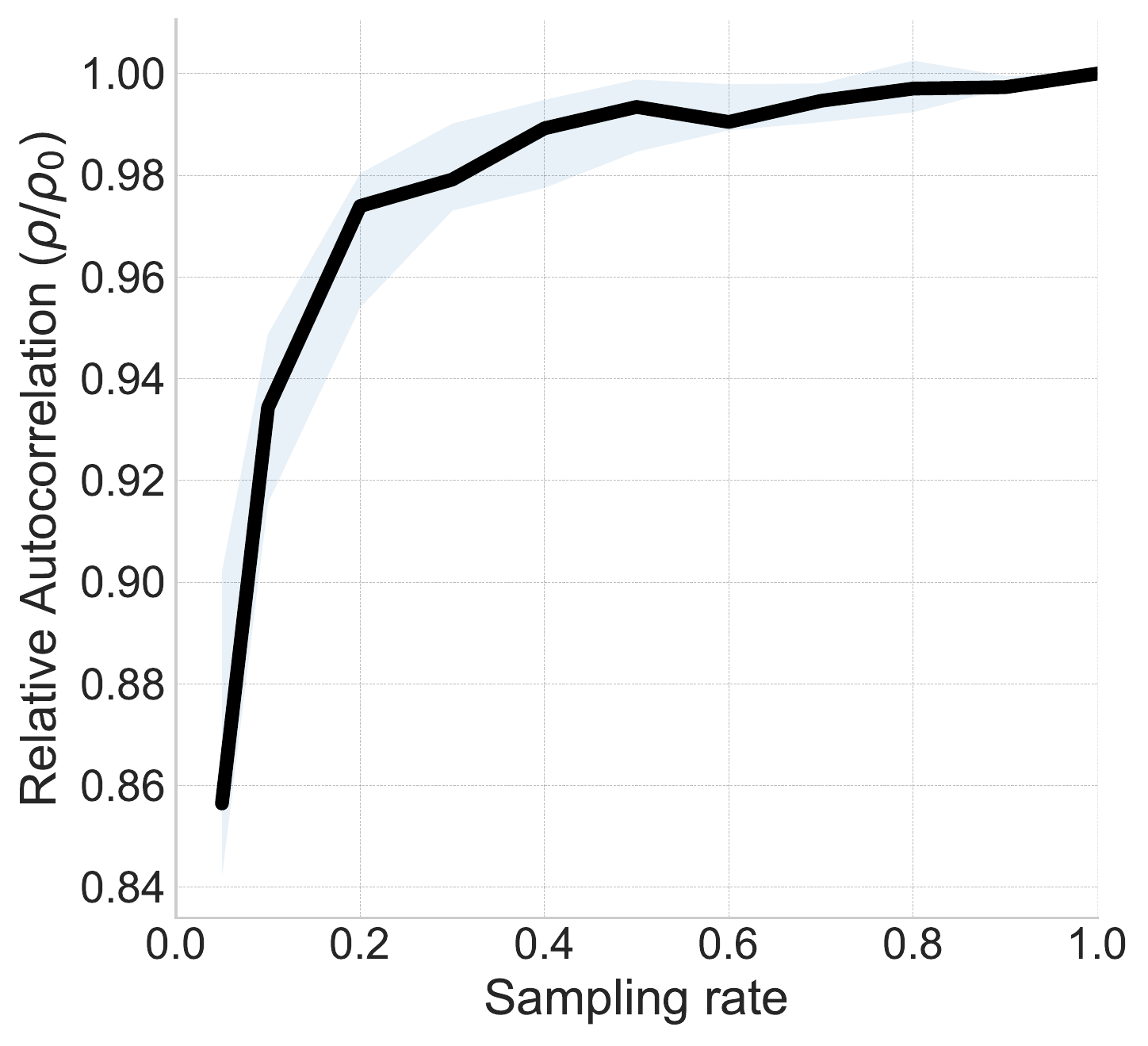}
\includegraphics[width=0.3\linewidth]{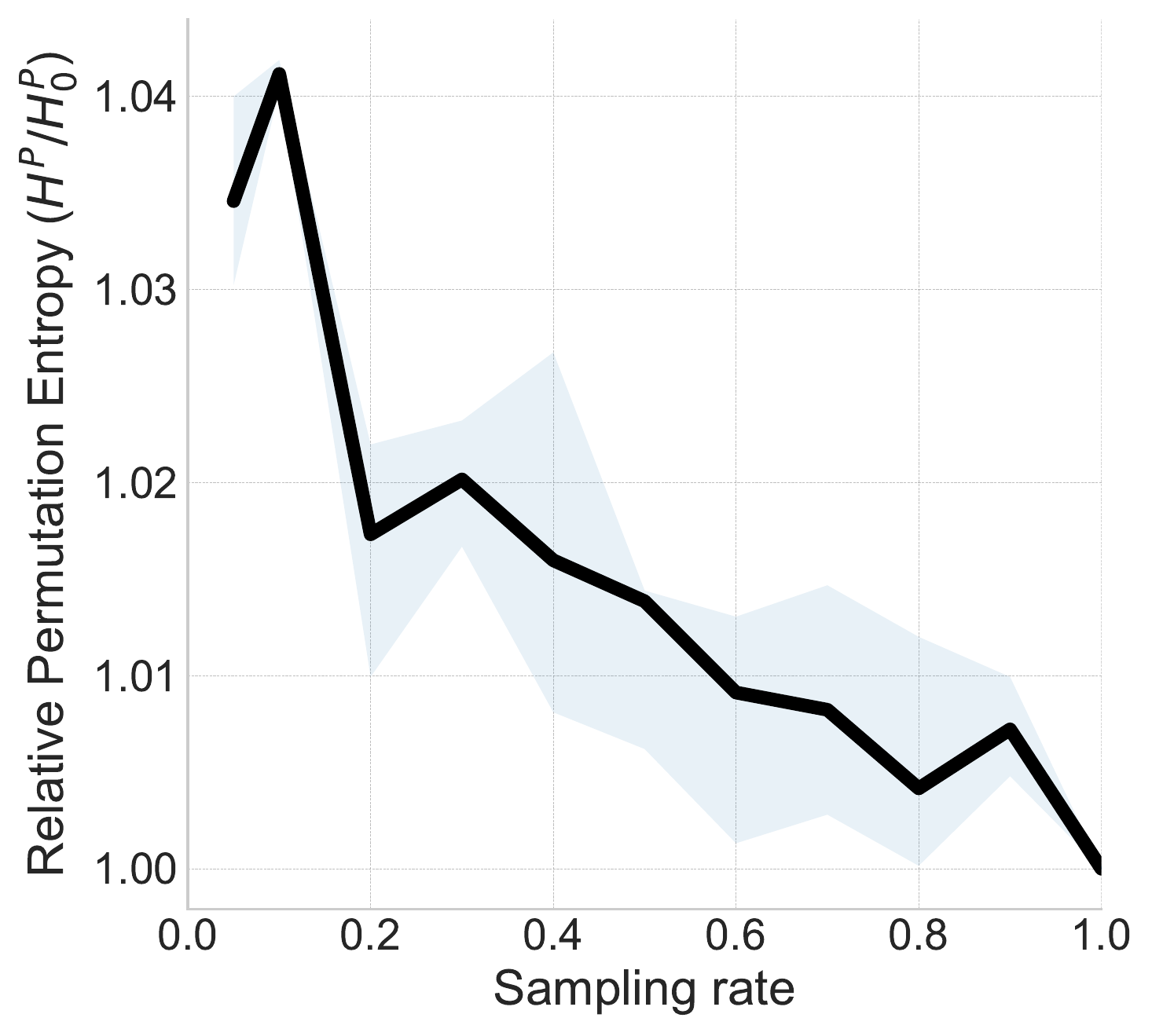}
\includegraphics[width=0.27\linewidth]{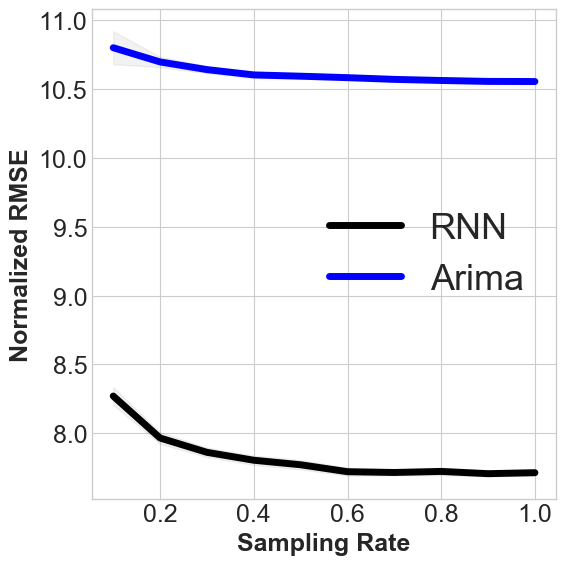}
\caption{Decay of predictability of (randomly) sampled Armstrong data. The (a) auto-correlation decreases at low sampling rates and (b) permutation entropy increases; and (c) and error of model-bases techniques increases}
\label{fig:armstrong-random}
\end{figure*}
\end{center}

\subsection{Measuring Predictability}
As measures of predictability we will use \emph{auto-correlation}, \emph{permutation entropy} and \emph{prediction error}.  The first two are model-free measures of predictability, while that latter requires model-based techniques such as ARIMA or RNN's (Recurrent Neural Network) to predict future values and compute their prediction error.\\

\emph{Auto-correlation} captures how well a time series is correlated (using Pearson correlation) with its own time-lagged versions, and has been widely used in finance~\cite{lim2013us}. Auto-correlation is also linked to the performance of auto-regressive linear models such as ARIMA, with higher auto-correlation leading to better performing auto-regressive models. For each time series $X$, we find the lag $\tau$, $1\leq \tau \leq 7$, which has the highest auto-correlation. We then use the found lag to compute the auto-correlation of all the sampled time series $Y$.   \\  

\emph{Permutation Entropy~\cite{Bandt2002permutation}} captures the complexity of a time series through statistics of its ordered sub-sequences (motifs), and it is used as a model-free, non-linear indicator of predictability, for example of infectious disease outbreaks~\cite{scarpino2019predictability} and human mobility~\cite{Song1018Science}. Permutation entropy, can be interpreted as the entropy of all the $d!$ possible motifs of fixed size $d$ present in a time series. The motifs represent ordinal patterns that measure the ordinal relation among successive time series values. As an example, if $x_1= 3, x_2 = 6, x_3 = 1$, then the ordinal pattern of this subsequence $\{x_1,x_2,x_3\}$ is $\phi(x_1,x_2,x_3) = (312)$ because $x_3\le x_1\le x_2$. Lower permutation entropy is associated with better predictability.
%\nazgol{Higher permutation entropy means lower predictability.}

\subsection{Forecasting Models}
For our model-based predictability measures, we use state-of-the-art forecasting models based on neural network architectures and autoregressive models.
The forecasting task is as follows. Given a daily time series describing cyber-attack events, we predict new events occurring in the future. Finally, we quantify the prediction error with the Root Mean Square Error (RMSE). Before feeding the data into the model we normalize the time series using z-scores by negating the values by their mean and dividing them by their standard deviation. In this way, the mean of the time series becomes $0$ and the standard deviation of $1$. Since the data is already normalized, we do not need to normalize the RMSE and thus, we can compare the RMSE of the same data at different samplings rates.

\subsubsection{Auto-regressive Models}
%\textbf{Negar to describe ARIMA model (see EFFECT paper)}
For our linear model, we use Auto Regressive Integrated Moving Average (ARIMA) to predict the future points of the time series. The ``AR'' part of the name indicates using the lagged observed time series as a regressor for predicting future values. The ``I'' part shows that this model differences the raw observations (subtracts each point in the observation from a previous time-step) to make it a stationary time series. The ``MA'' part indicates that the model uses a linear combination of lags of the forecast errors as the regression error. An ARIMA model is specified as ARIMA(p,d,q) in which p is the number of autoregressive terms, d is the number of differences applied to make the time series stationary, and q is the order of moving average for the forecast errors. We performed a grid search to choose the set of parameters that minimizes the AIC (Akaike Information Criterion) value of goodness-of-fit.

\begin{center}
\begin{figure*}[t]
\centering
\includegraphics[width=0.3\linewidth]{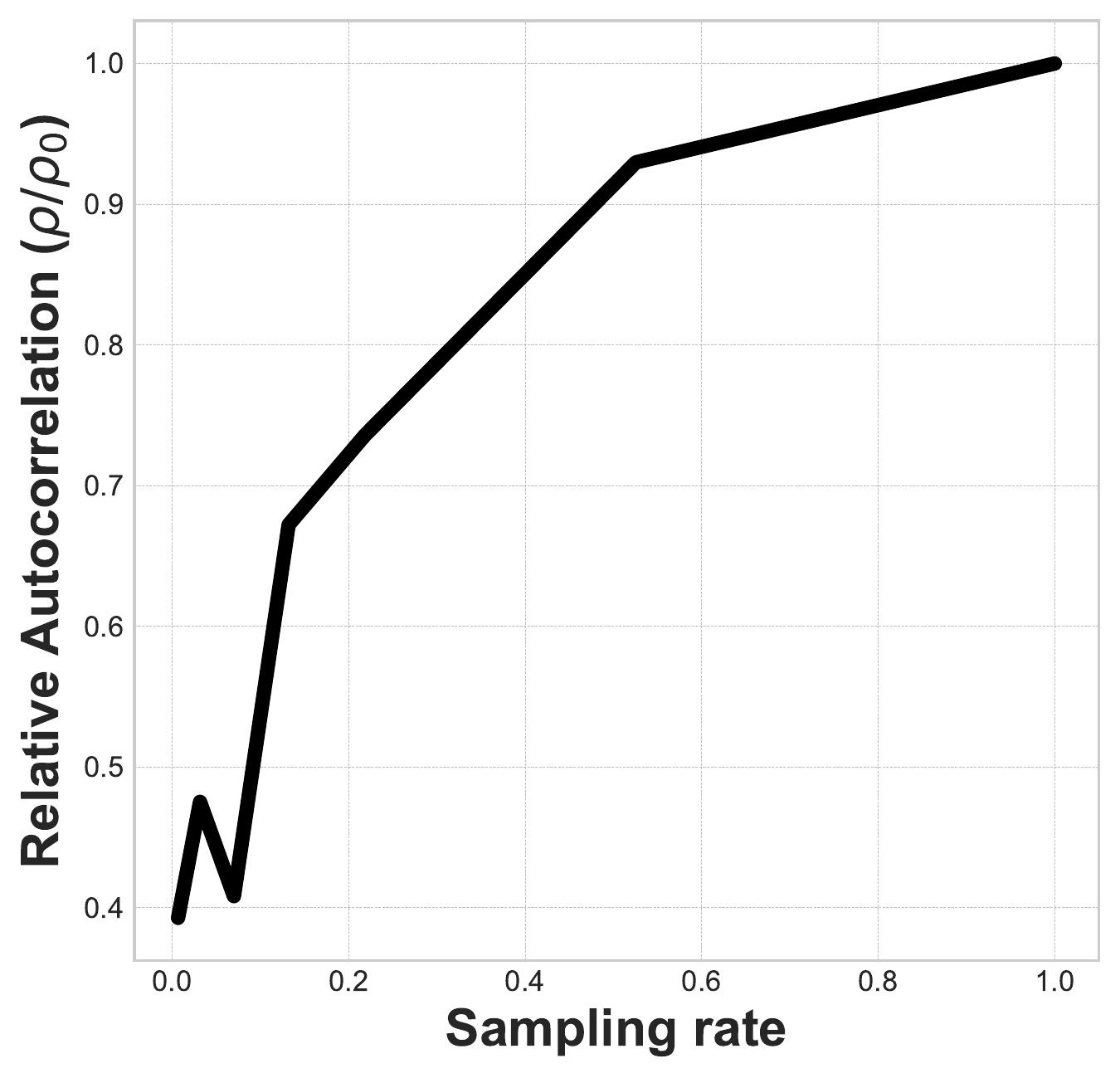}
\includegraphics[width=0.3\linewidth]{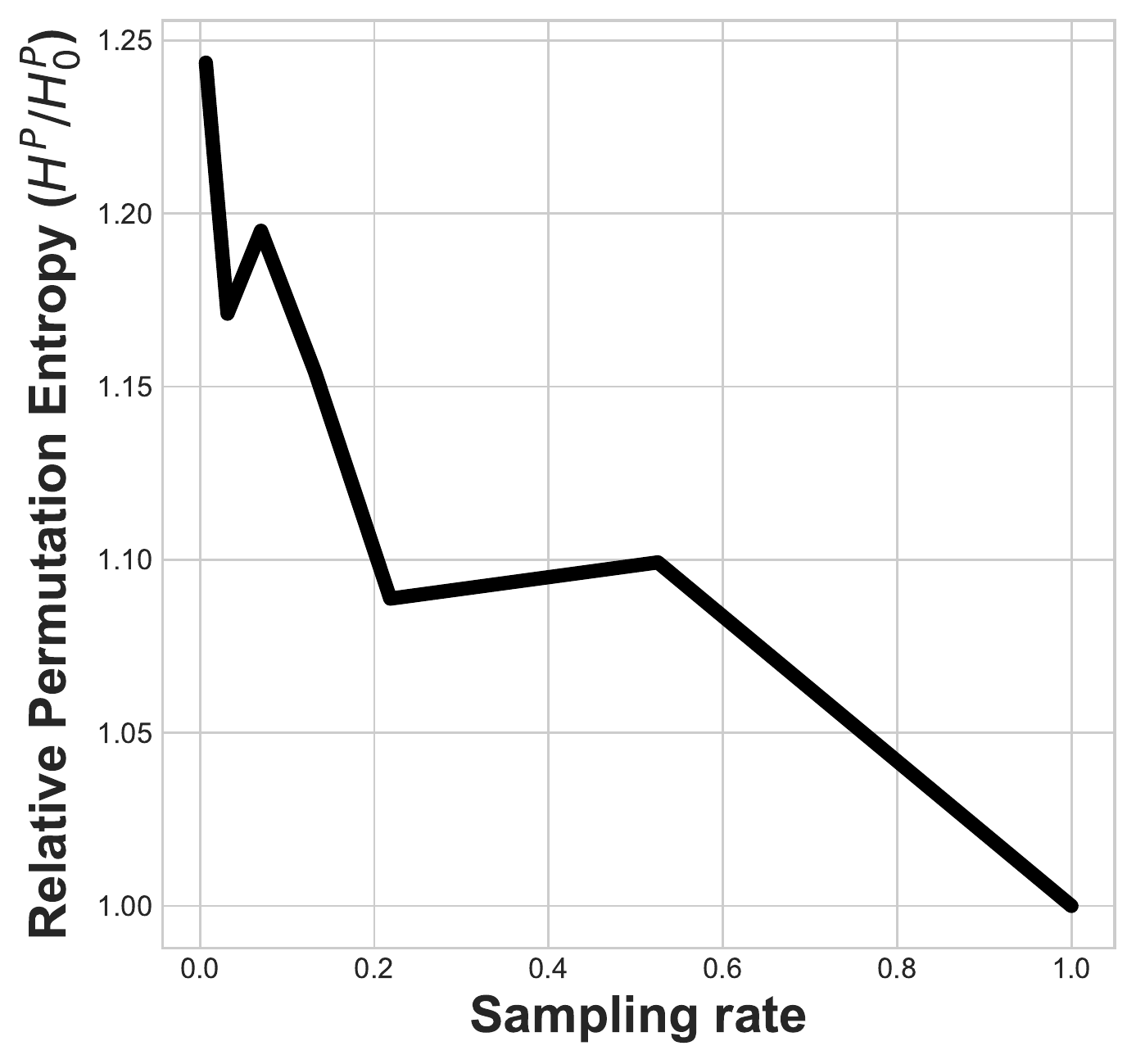}
\includegraphics[width=0.27\linewidth]{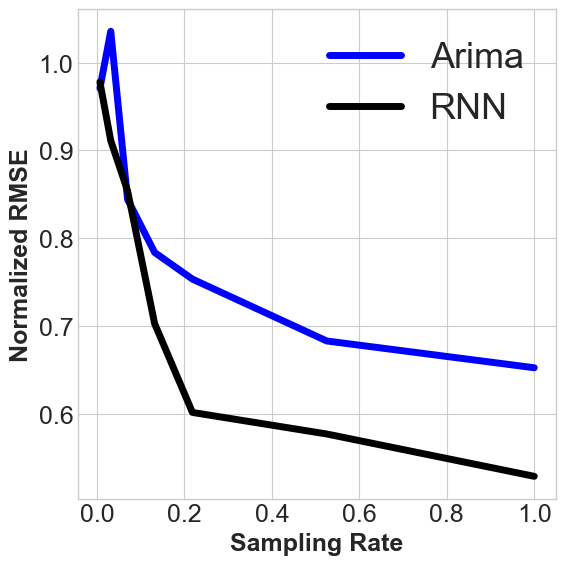}
\caption{Decay of predictability of ISI data under real-world sampling. The (a) auto-correlation decreases at low sampling rates; (b) permutation entropy increases; and (c) and error of model-bases techniques increases.
% \andres{the scale on the Y-axis should be on the same scale as the random sampling in figure \ref{fig:isi-random}  }
}
\label{fig:isi-realworld}
\end{figure*}
\end{center}
\begin{center}
\begin{figure*}[t]
\centering
\includegraphics[width=0.3\linewidth]{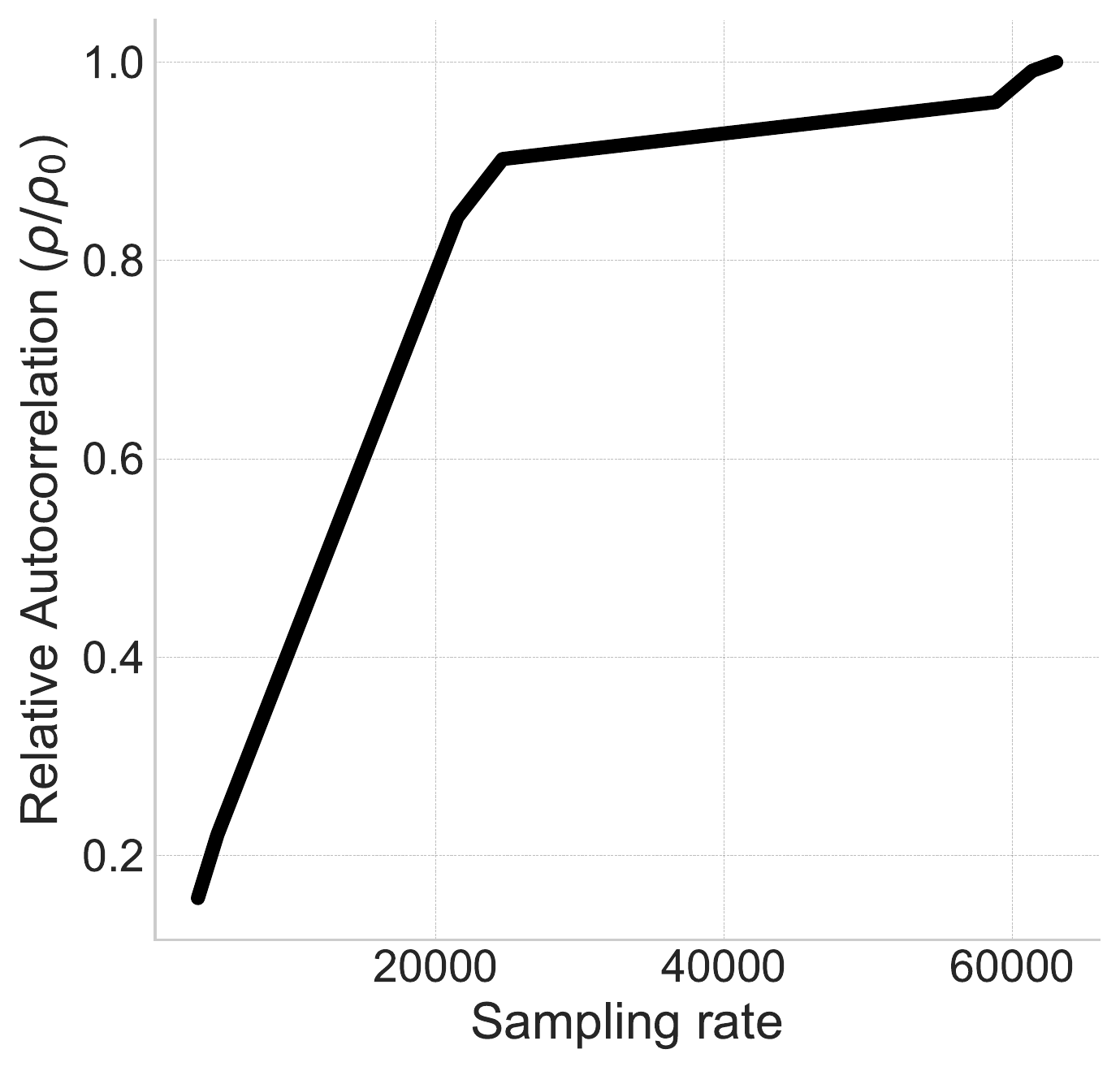}
\includegraphics[width=0.3\linewidth]{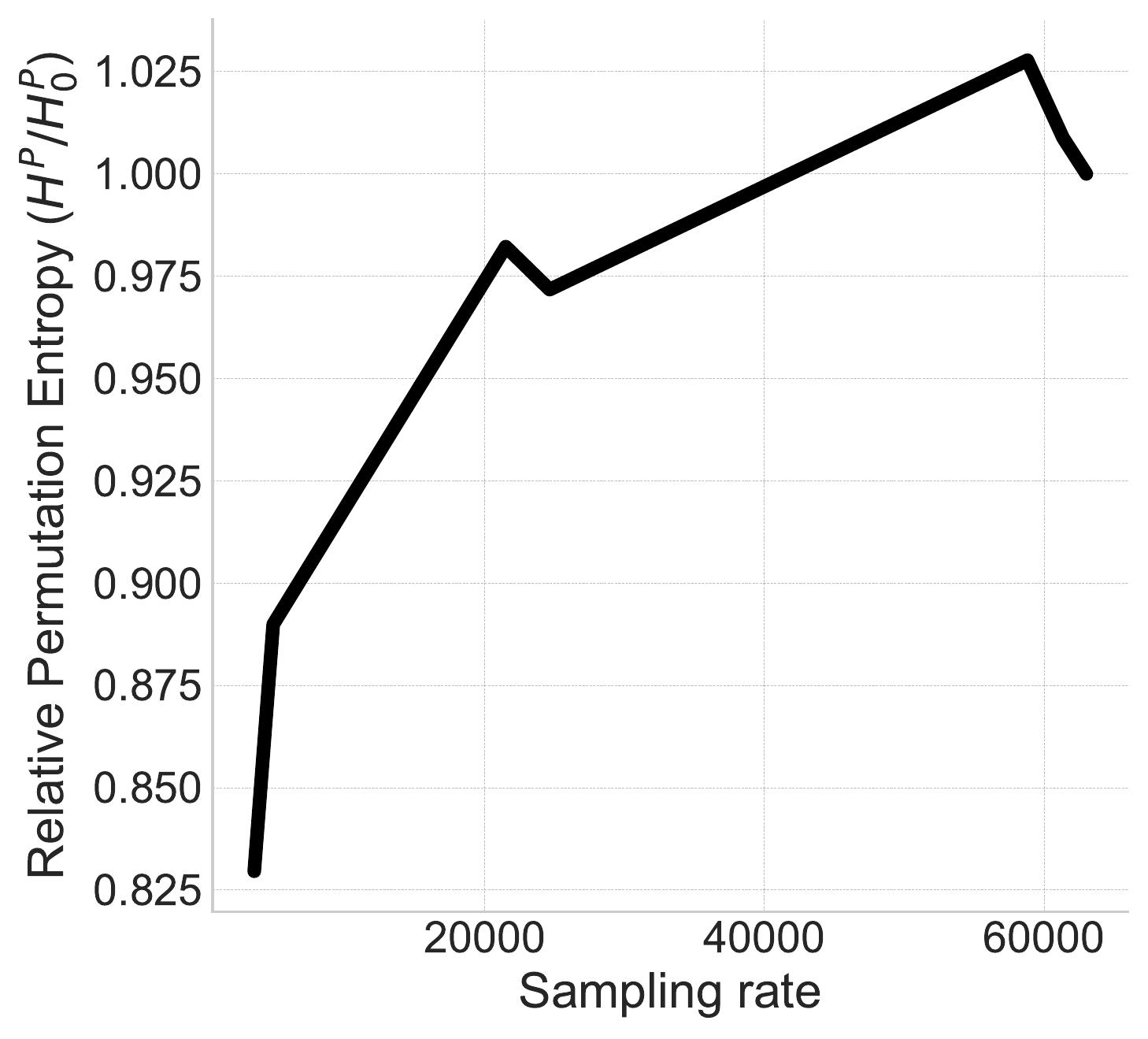}
\includegraphics[width=0.27\linewidth]{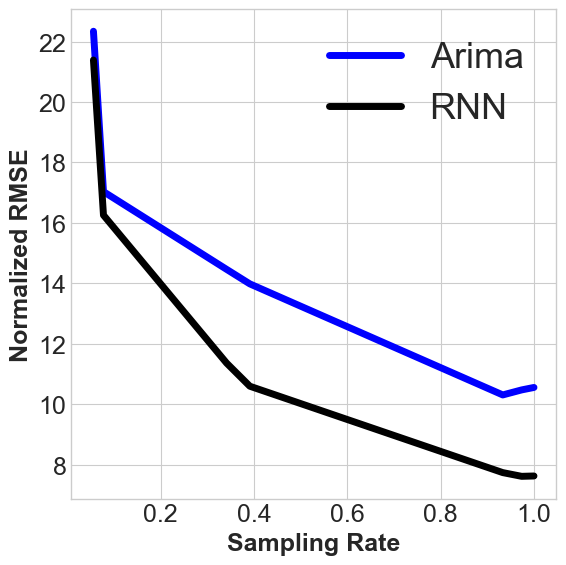}
\caption{Decay of predictability of Armstrong data under real-world sampling for three out of the four measures. The (a) auto-correlation decreases at low sampling rates; (b) permutation entropy decreases; and (c) and error of model-bases techniques increases.}
\label{fig:armstrong-realworld}
\end{figure*}
\end{center}

\subsubsection{Neural Networks}
Neural network models have grown dramatically in popularity across many applications, including forecasting temporal phenomena. Neural network-based models can capture non-linearities by using multiple layers of non-linear activation functions. The caveat is that such models typically require a large amount of training data to accurately estimate the model parameters.
Many variants of recurrent neural network units have been proposed including Long Short Term Memory
(LSTM)\cite{hochreiter1997long} which has been widely used for time series prediction \cite{qin2017dual} and even more specifically for cyber attack prediction \cite{aditham2017lstm}. Each LSTM unit contains cell state plus input, output and forget gates. This architecture is designed to remember long term dependencies and forget irrelevant information which can be very useful for time series prediction. We use a simple one layer LSTM plus activation function for training the model. Since we had limited amount of data more complex models failed due to overfitting.%We used the same LSTM architecture and training method as in Goyal et. al~\cite{goyal2018discovering}.

%%% REMOVE?
%For this work we look at \nazgol{malicious} emails targeted towards 2 organizations, Armstrong and ISI. We also include spam email and graymail. Graymail refers to emails that fall between spam and welcomed email where some users value and others prefer to block \cite{sculley2008filtering}. 

%Some of these emails were blocked by security appliances of companies and some others were delivered, marking a successful cyber attack. In this section we will describe the datasets in more detail.

\section{Results}

In this section, we show our findings surrounding the effect of sampling/filtering on the predictability of cyberattacks. We use the two data sets described in previous sections, ISI and Armstrong.  For each data set, we use both random sampling and real-world sampling to compare their effects on predictability. See Methods Section~\ref{sec:sampling} for details on the different sampling approaches.  

We measure predictability with auto-correlation, permutation entropy and prediction error. For prediction error, we use ARIMA as a representative for linear models and LSTM as a representative for non-linear models. We use RMSE (Root Mean Squared Error) to measure prediction error. 

%In all of the experiments below, we first trained the models on the first month to predict the next they, then added the value of that day to historical data to predict the 
% {\color{red} Nazgol not clear": What was used for training and what for prediction; how you did the moving window ,etc}
In all of the experiments below, for each model, data set, and sampling rate, we first trained the models with the first month of the historical data. Then, we performed next day predictions on the rest of the data set. The models were retrained with the whole historical data every week of predictions. 
ARIMA generally uses all the historical data as input to predict, however, for the RNNs we only used the 7 previous days as input. 

%For real world sampling we use the percentage of data remained after each filter on the X axis instead of the actual thresholds of filters.

%\andres{where/what are figures mentioned below?} 

%\subsection{Loss of Predictability}

\subsection{Random Sampling}
We first examine the loss of predictability due to sampling under the random sampling scenario, where the email events have equal, but variable probability to be filtered by the security appliance. Figures~\ref{fig:isi-random}  and \ref{fig:armstrong-random} show that the predictability decays at lower sampling rates for both our data sets. This effect holds true no matter the measure of predictability used. Whether it is a model-free measure (such as auto-correlation or permutation entropy)
or using our model-based measures (prediction error).
Research question RQ1 asked whether non-linear methods are more or less robust to the effects of sampling. Comparing the error of the ARIMA vs RNN predictions in both enterprises we highlight two main observations: 
\begin{itemize}
    \item The RNN (non-linear) method clearly outperforms the linear based ARIMA predictions, at all sampling rates, in both enterprises.
    \item For ISI data (Figure ~\ref{fig:isi-random}), the accuracy of the predictions deteriorates more sharply and abruptly at lower sampling rates (around 40\%) compared to the RNN. Whereas for Armstrong (Figure~\ref{fig:armstrong-random}), both methods exhibit similar and milder behavior as a function of the sampling rate.
\end{itemize}

\begin{figure}[t]
\centering
  \begin{subfigure}[b]{0.47\textwidth}
  \centering
    \includegraphics[width=\textwidth]{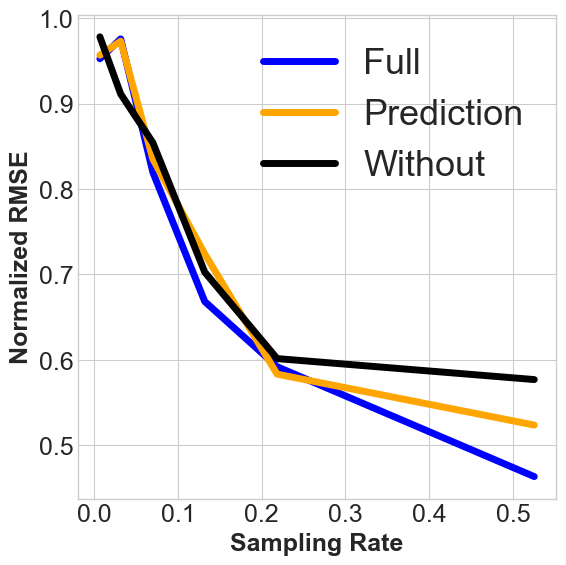}
    \caption{ISI data (with real-world sampling).}
    \label{fig:isi_ext}
  \end{subfigure}%
   \hspace{1em}%
 \centering
  \begin{subfigure}[b]{0.47\textwidth}
  \centering
    \includegraphics[width=\textwidth]{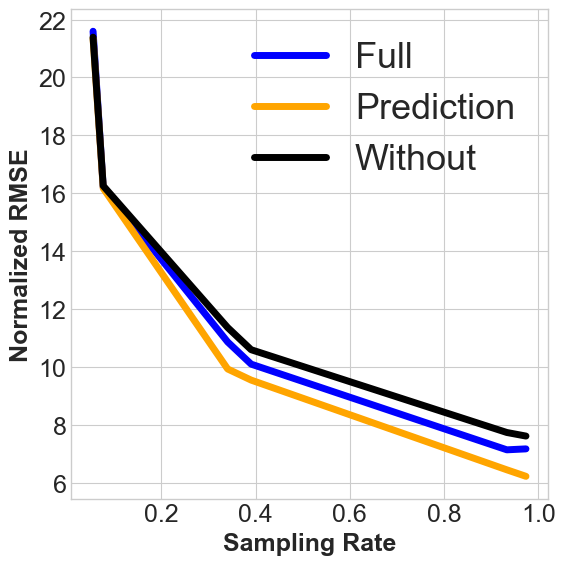}
\caption{Armstrong data (with real-world sampling)}
\label{fig:arm_ext}
  \end{subfigure}
  \caption{Using raw (unfiltered) data as external signal to predict the data with real-world sampling.  Full: unfiltered data as external signal; Prediction: predicted unfiltered data as external signal; Without: without external signal}
\end{figure}

\subsection{Real-World Sampling}
% \andres{Mostly a note for a future version: I would put the equivalent of the sampling rate (i.e., percentage of data) in the x-axis of figures \ref{fig:isi-realworld} and  \ref{fig:armstrong-realworld} .  }
Under the real-world sampling scenario, predictability also decays as more of the unsolicited emails are removed by the security appliances. 
Figures~\ref{fig:isi-realworld} and \ref{fig:armstrong-realworld} show the loss of predictability under real-world sampling scenarios. Note that the curves look similar to those for random sampling, with the exception of Armstrong's decreasing permutation entropy trend. Yet the effects are much more pronounced, which is evident in the fast incline of error for both ARIMA and RNN methods alike. Addressing question RQ1, we highlight that,
\begin{itemize}
    \item Real-world sampling affects similarly to both linear and non-linear methods. Specifically, for low sampling rates, both methods have a similar high error. 
\end{itemize}
Regarding question RQ2, we compare real-world sampling vs random sampling.
\begin{itemize}
    \item Our results suggest that in real-world scenarios, partial observability has a higher impact on predictability.
\end{itemize}
Together, both results suggest that the practical implications of incomplete data when predicting cyber-attacks not only are significant, but are stronger than what the current theory indicates. We have empirically validated that our models and theories do assert an important limit in predictability. Yet, as our results show, further steps and future work are needed to better quantify, mitigate and predict the effects of partial data in real-world situations.

\subsection{Effects of External Signals}
Results presented above show how predicting cyber-attacks become increasingly difficult as we consider more dangerous threats (which are fewer in number). At the same time predicting attempted cyber-attacks, which are of less interest, is considerably easier. We can take advantage of this fact and use the prediction of the attempted cyber-attacks as an external signal.
%Since forecasting unfiltered signal is easier, we can use the predictions as an external signal. 
We test this theory using the RNN model, since it consistently gave better results than ARIMA. We first forecast the external signal for the next day ($\hat{x}_{t + 1}$), then feed 7 historical days of both target data and external data plus $\hat{x}_{t + 1}$, to the model. We ran this experiment at different sampling rates (except for the raw data). 

Figures \ref{fig:arm_ext} and \ref{fig:isi_ext} show results for Armstrong and ISI data, both with real-world sampling. In both plots, "Prediction" presents the experiment described above. To compare the results, we also plot predictions without external signals, the same results presented in Figures \ref{fig:armstrong-realworld}C and \ref{fig:isi-realworld}C, as "Without". The "Full" line represents the error of forecasts using the unfiltered data as an external signal. We expected "Full" and "Without" to be upper bounds and lower bounds for "Prediction" respectively. This is the case in Figure \ref{fig:isi_ext}; however, in Figure \ref{fig:arm_ext} "Prediction" outperforms "Full". This could be because of the high permutation entropy of unfiltered Armstrong which also causes an unexpected trend in Figure \ref{fig:armstrong-realworld}B. %Indeed, the error in Armstrong is higher that for ISI's data.
% \nazgol{Andres do you agree?yes!}. 
Figures \ref{fig:isi_ext} and \ref{fig:arm_ext} also show that the raw signal is a better external signal when we are trying to predict signals with similar sampling rates. This suggests that signals with higher yet similar sampling rates could be the best candidates for external signals.  % data  raises the question that whether using 

\section{Conclusion}
Artificial intelligence and machine learning have generated much excitement in recent years with their ability to identify elusive patterns in large volumes of data. The hope for the cybersafety community is that AI systems can be trained to recognize precursors of malicious events, such as cyber-attacks, in the voluminous data streams from open online sources. Our work identifies inherent obstacles to realizing this vision. Importantly, the performance of AI systems is only as good as the training data they are provided. However, if training data represents only a partial observation of the processes of interest, the performance of the predictive model necessarily degrades. In the cybersecurity setting, this means that AI systems trained on filtered data -- e.g., successful cyber-attacks or events passing through an organization's firewall -- will not be able to accurately predict future cyber-attacks. Using data from two organizations, we demonstrated the loss of predictability as more and more of the data was filtered, for example, by the organization's security appliances.%, showing how predicting cyberattacks becomes increasingly difficult as we consider more dangerous threats. %Also they both follow a curve pattern (which is a popular pattern in our results). These curves represent how predictability converges to its optimal value with high sampling rates and brings up the question that whether these curves could be used to give a lower bound on predictability when we don't observe the whole data.  %This suggests that in order to improve predictability, researchers should be working with as complete a stream of ground truth data as possible.

Our work identifies potentially fruitful avenues for future research. Figures~\ref{fig:armstrong-realworld} and \ref{fig:isi-realworld} show that non-linear models such as RNN do have an advantage over linear models, however, this gap decreases for lower sampling rates. Suggesting that in practice, the full advantage of these methods is not fully exploited.
Future work will be devoted to further increase this gap, investigating the possibility of mitigating the loss of information using new non-linear, neural network based methods. 
Second, our framework is very flexible and general, and can be applied to a myriad of applications beyond cybersecurity, for example, predicting armed conflict, protest, and political unrest around the world. 

\section*{Acknowledgements}
This work was supported by the Office of the \textit{Director of National Intelligence} (ODNI) and the \textit{Intelligence Advanced Research Projects Activity} (IARPA) via the \textit{Air Force Research Laboratory} (AFRL) contract number FA8750-16-C- 0112, and by the \textit{Defense Advanced Research Projects Agency} (DARPA), contract number W911NF-17-C-0094. The U.S. Government is authorized to reproduce and distribute reprints for Governmental purposes notwithstanding any copyright annotation thereon. Disclaimer: The views and conclusions contained herein are those of the authors and should not be interpreted as necessarily representing the official policies or endorsements, either expressed or implied, of ODNI, IARPA, AFRL, DARPA, or the U.S. Government.
% \end{acks}

% \newpage
\bibliographystyle{unsrt}  
\bibliography{references}

\end{document}